\documentclass[aps,superscriptaddress,onecolumn,a4paper,reprint]{revtex4}
\usepackage[colorlinks=true, pdfstartview=FitV, linkcolor=blue, citecolor=red, urlcolor=black]{hyperref}
\usepackage{graphicx}
\usepackage[all]{xy}
\usepackage{amsmath}
\usepackage{amssymb}
\usepackage{tensor}
\newcommand{\be}{\begin{equation}}
\newcommand{\ee}{\end{equation}}
\newcommand{\ben}{\begin{eqnarray}}
\newcommand{\een}{\end{eqnarray}}
\newcommand{\bes}{\begin{subequations}}
\newcommand{\ees}{\end{subequations}}
\def\bal#1\eal{\begin{align}#1\end{align}}

\newcommand{\bfi}{\begin{figure}}
\newcommand{\efi}{\end{figure}}
\newcommand{\bc}{\begin{center}}
\newcommand{\ec}{\end{center}}

\newcommand{\sgn}{\mbox{sgn}}

\newcommand{\sech}{\mbox{sech}}

\newcommand{\LL}{{\cal L}}

\newcommand{\p}{\partial}

\begin{document}

\title{Kinks in cuscuton-like models with two scalar fields}
\author{I. Andrade}\affiliation{Departamento de Licenciaturas em Ci\^encias Naturais/Biologia, Universidade Federal do Maranh\~ao, 65400-000 Cod\'o, MA, Brazil}
\author{R. Menezes}\affiliation{Departamento de Ci\^encias Exatas, Universidade Federal
da Para\'{\i}ba, 58297-000 Rio Tinto, PB, Brazil}


\begin{abstract}
This work deals with the presence of localized structures in relativistic systems described by two real scalar fields in two-dimensional spacetime. We consider the usual two-field model with the inclusion of the cuscuton term, which couples the fields regardless the potential. First we follow the steps of previous work to show that the system supports a first-order framework, allowing us to obtain the energy of solutions without knowing their explicit form. The cuscuton term brings versatility into the first-order equations, which gives rise to interesting modifications in the profiles of topological configurations, such as the smooth control over their slope and the internal structure of the energy density.

\end{abstract} 

\maketitle

\section{Introduction}
Topological defects are of interest in physics in general, playing an important role in formation of structures in the early Universe and have been studied in a diversity of scenarios involving the cosmic evolution and other issues in high energy physics \cite{rajaraman, Wilets,Vilenkin} and several areas of non-linear science \cite{Whitham, Eschenfelder, Davidov, Murray, Walgraef}. Kinks are solutions of the equations of motion that connect two minima of a potential and arise in models with one or more fields in (1,1) spacetime dimensions.  Due to their simplicity, kinks are often used to describe specific behaviors in many physical situations, see the Refs.~\cite{Toharia,Bischof,Leon,Martin}. 

Recent investigations of scalar fields with non-canonical Lagrangians have also been considered to generate topological structures. These models are motivated by the current studies in cosmology. In particular, kinematic modifications of scalar fields have been proposed as an attempt to explain the accelerated expansion of the Universe \cite{in1}, and as possible solutions to cosmic problems \cite{co1,co2}. A special non-dynamical class of generalized models is the cuscuton one, which was introduced as a field with infinite speed of propagation of sound \cite{cuscuton1,cuscuton2}. The cuscuton does not possess dynamical degrees of freedom associated with its temporal evolution, as it does not contribute to temporal derivative terms in the equation of motion. There have been several papers published over the years that discuss the cuscuton term, as referenced in \cite{cuscuton3,cuscuton4,cuscuton5,cuscuton6,cuscuton7,cuscuton8}.

{ For topological and non-topological defects, the presence of non-linear interactions in the dynamics of scalar field models has been studied in Refs.~\cite{Babichev:2006cy,trilogia1,Adam:2007ag,Bazeia:2016xrf,Babichev:2007tn,Adam:2008rf,Casana:2015cla,Atmaja:2015umo,Bazeia:2017nas}, within the context of discrete (kinks) and continuous (Q-balls, vortices, and monopoles) symmetries. The possibility of such extensions of canonical models leads to interesting features in the solutions and their properties. For instance, a distinct character that may appear in non-canonical models is the twinlike character. In Ref.~\cite{Andrews:2010eh}, it was introduced a generalized model that engender a mass parameter which supports the same solutions and energy density of the standard model; see also Ref.~\cite{Bazeia:2011sb}. In this scenario, the generalized and standard models are called twins. Notwithstanding that, they can be distinguished by their linear stability, whose properties differ from one another.}

In Ref.~\cite{cuscutonigor}, the cuscuton term was added to the canonical kinematics in the Lagrangian density of a single real scalar field, 
\be
\LL = \frac12\p_\mu\phi\p^\mu\phi +\frac{f(\phi)\p_\mu\phi\p^\mu\phi}{\sqrt{|\p_\mu\phi\p^\mu\phi|}} -V(\phi),
\ee
{ with $f(\phi)$ and $V(\phi)$ being arbitrary functions of the field. It is possible to rewrite the middle term in the form $f(\phi)\,\sgn(\p_\mu\phi\p^\mu\phi)\sqrt{|\p_\mu\phi\p^\mu\phi|}$, where $\sgn(z)$ is the signal function with argument $z$}. The properties of this model were studied and kink solutions for the static case ($\phi_s(x)$) were obtained analytically. { The way the additional term was introduced leads to two characteristics that we emphasize}: i) the cuscuton term does not contribute to the equation of motion; ii) although it contributes to the linear stability equation, it does not appear in terms involving derivatives of the perturbation $\eta$, where $\phi(x,t)=\phi_s(x)+\eta(x,t)$. Here, we will study how the presence of another scalar field affects these features.

The paper is organized as follows. In Sec.~\ref{sec2} we introduce a new class of non-canonical models and analyze their properties. We also investigate the first-order formalism and explore the peculiarities arising from the inclusion of a new term, as well as the constraints considered.
Next, in Sec.~\ref{sec3}, we provide three explicit examples in detail to illustrate the procedure. Kink profiles and their energy densities were obtained analytically. We conclude our work in Sec.~\ref{sec4}, where we comment on the main results and discuss possible perspectives. For simplicity, we consider all quantities to be dimensionless and set $c=\hbar=1$. 

\section{The model}\label{sec2}
We start by considering a Lagrangian density for two real scalar fields, $\phi$ and $\chi$, in the two-dimensional $(1,1)$ Minkowski spacetime, with $\eta_{\mu\nu}={\rm diag}(+,-)$,
\be\label{model}
\LL = \frac12\p_\mu\phi\p^\mu\phi +\frac12\p_\mu\chi\p^\mu\chi +\frac{f(\chi)\p_\mu\phi\p^\mu\phi}{\sqrt{|\p_\mu\phi\p^\mu\phi|}} -V(\phi,\chi).
\ee
This model is a generalization for two fields of the one considered in Ref.~\cite{cuscutonigor}. In addition to the interaction between the fields $\phi$ and $\chi$ through the potential term, there is a non-standard coupling between the fields in which the function $f(\chi)$ plays the role of a coupling parameter for the cuscuton term.

The equations of motion derived from the above model are
\be\label{eom}
\p_\mu\left(\left(1 +\frac{f(\chi)}{\sqrt{|\p_\nu\phi\p^\nu\phi|}}\right)\p^\mu\phi\right) = -V_\phi  \,\,\,\,\,\,\,\, {\rm and} \,\,\,\,\,\,\,\, \Box\chi = \frac{f_\chi\p_\mu\phi\p^\mu\phi}{\sqrt{|\p_\mu\phi\p^\mu\phi|}} -V_\chi,
\ee
where we use the notation $V_\phi=\p V/\p\phi$, $V_\chi=\p V/\p\chi$ and $f_\chi=df/d\chi$.
The energy-momentum tensor $T_{\mu\nu}$ for this generalized model is given by
\be\label{Tmn}
T_{\mu\nu} = \left(1 +\frac{f(\chi)}{\sqrt{|\p_\lambda\phi\p^\lambda\phi|}}\right)\p_\mu\phi \p_\nu\phi + \p_\mu\chi\p_\nu\chi -\eta_{\mu\nu}\left(\frac12\p_\mu\phi\p^\mu\phi +\frac12\p_\mu\chi\p^\mu\chi +\frac{f(\chi)\p_\mu\phi\p^\mu\phi}{\sqrt{|\p_\mu\phi\p^\mu\phi|}} -V(\phi,\chi)\right).
\ee
Following Ref.~\cite{trilogia1}, we constrain ourselves to obey the null energy condition (NEC), that imposes $T_{\mu \nu} n^\mu n^\nu \geq 0 $, where $n^\mu$ is an arbitrary null vector ($\eta_{\mu\nu}n^\mu n^\nu=0$). For the above energy-momentum tensor, this condition leads to
\be\label{NEC}
\left(\left(1 +\frac{f(\chi)}{\sqrt{|\p_\lambda\phi\p^\lambda\phi|}}\right)\p_\mu\phi\p_\nu\phi + \p_\mu\chi\p_\nu\chi\right)n^\mu n^\nu \geq 0,
\ee
that depends on $\chi$ and the derivative of both fields.

Since we are interested in topological structures, we consider static solutions $\phi=\phi(x)$ and $\chi=\chi(x)$, with boundary conditions $\phi(x\to\pm\infty)=v_\pm$ and $\chi(x\to\pm\infty)=w_\pm$, where $v_\pm$ and $w_\pm$ are constants that represent the asymptotic values of the fields $\phi$ and $\chi$. In this situation, the equations of motion \eqref{eom} reduce to
\bes\label{seom}
\bal
\label{eomP}\phi''&= V_\phi -\frac{f_\chi\chi^\prime\phi^\prime}{|\phi^\prime|},\\
\label{eomC}\chi''&= V_\chi +f_\chi|\phi^\prime|,
\eal
\ees
Here, the prime denotes derivative with respect to $x$. In contrast to the one-field model investigated in Ref.~\cite{cuscutonigor}, the cuscuton term contributes to the equations above due to the presence of the function $f(\chi)$. However, this contribution is of first order in the derivative of the fields, and there are no second-order terms. This is one of the characteristics of a Cuscuton term.

The non-zero components of the energy-momentum tensor in Eq.~\eqref{Tmn} are given by
\bes 
\bal 
\label{rho}
&\rho = \frac12{\phi^\prime}^2 +\frac12{\chi^\prime}^2 +f(\chi)|\phi^\prime| +V(\phi,\chi),\\
&\tau = \frac12{\phi^\prime}^2 +\frac12{\chi^\prime}^2 -V(\phi,\chi).
\eal 
\ees
The energy and stress densities are denoted by $\rho$ and $\tau$, respectively. To ensure that the energy $E=\int^\infty_{-\infty} dx \rho$ is finite, the function $f(\chi)$ must be chosen properly. The component related to the stress of the solution does not depend on the function $f(\chi)$. 

Multiplying the Eq.~\eqref{eomP} by $\phi^\prime$ and the Eq.~\eqref{eomC} by $\chi^\prime$, it is possible to integrate the sum of them to obtain
\be\label{vinc}
\frac12{\phi^\prime}^2 +\frac12{\chi^\prime}^2 = V(\phi,\chi),
\ee
where we have taken the constant of integration as zero to obtain the stressless solutions ($\tau=0$) with finite energy \cite{trilogia1,trilogia2}. In contrast to the one field situation \cite{trilogia1}, the stressless condition is not sufficient to obtain a first order formalism here. 
Substituting the stressless constraint \eqref{vinc} in the energy density \eqref{rho}, we get
\be\label{rho0}
\rho(x) = \left(1 +\frac{f(\chi)}{|\phi^\prime|}\right){\phi^\prime}^2 +{\chi^\prime}^2.
\ee
The condition imposed in Eq.~\eqref{NEC} ensures the positivity of energy density. Following the lines of Ref.~\cite{trilogia2}, we construct a first order formalism by introducing the new functions $P(\phi)$ e $Q(\chi)$ as
\bes\label{fo}
\bal
\label{foP}\phi^\prime &= \pm\left(1+\frac{f(\chi)}{|\phi^\prime|}\right)^{-1}P_\phi,\\
\chi^\prime &= \pm Q_\chi  \label{foQ}.
\eal
\ees
We rewrite the energy density \eqref{rho0} in the form
\be\label{rhoW}
\rho(x) = \rho_1(x)+\rho_2(x),\quad  {\rm with} \quad \rho_1(x) = \pm\frac{dP}{dx}  \quad\text{and}\quad \rho_2(x)=\pm\frac{dQ}{dx}.
\ee
The signals in Eqs.~\eqref{fo} and \eqref{rhoW} are independent and must be chosen based on the values of the functions $P(\phi)$ and $Q(\chi)$ for the boundary conditions $\phi(x\to\pm\infty)=v_\pm$ and $\chi(x\to\pm\infty)=w_\pm$.  Thus the energy can be expressed as the variation of $P$ and $Q$ and we can show that $E = E_1 +E_2$, where $E_1=|P(v_+)-P(v_-)|$ and $E_2=|Q(w_+)-Q(w_-)|$.  It is possible to calculate the energy without knowing the solutions, depending only on their respective boundary conditions in the functions $P$ and $Q$. Finally, we note that the function of the cuscuton term does not contribute to the energy.

To simplify the system of Eqs.~\eqref{fo}, we focus our attention on monotonic topological configurations. If $P$ and $Q$ are chosen to be monotonically non-decreasing functions  and $f(\chi)$ as an even function, these equations become
\bes\label{fokk}
\bal
\label{foPkk}\phi^\prime & = P_\phi -f(\chi),\\
\label{foQkk}\chi^\prime &=  Q_\chi.
\eal
\ees
These equations hold for the kink-kink situation ($\phi_k^\prime$ and $\chi_k^\prime$ non-negative), since $P_\phi \geq f(\chi)$.  Other situations can be obtained by changing the sign of $x\to-x$ in the above first order equations. For instance, to obtain the pair kink-antikink we change the \eqref{foQkk} for $-\chi^\prime =  Q_\chi$, keeping the other equation unchanged; and so on. 

From the stressless constraint in Eq.~\eqref{vinc}, we can write the potential in the form
\be
V(\phi,\chi) = \frac12{(P_\phi -f(\chi))}^2 +\frac12Q_\chi^2.
\ee
It is clear to see that the above potential is non-negative, i.e, $V(\phi,\chi)\geq0$. Topological solutions connect potential minima, which correspond to constant values $(\phi,\chi)=(v_i,w_i)$ such that $V(v_i,w_i)=0$. As we focus on kink-kink configurations, we study the cases where $v_+>v_-$ and $w_+>w_-$, and seek solutions that connect the minima $(v_-,w_-)$ to $(v_+,w_+)$. 

As the Eq.~\eqref{foQkk} does not depend of $\phi$,  we can obtain the solution $\chi(x)$ by evaluating the integral
\be\label{eqC}
x-x_0 = \int^\chi \frac{ d\tilde{\chi}}{Q_{\tilde{\chi}}} = h(\chi) \quad\quad \text{to get} \quad\quad \chi(x) = h^{-1}(x-x_0),
\ee 
for a specific choice of $Q(\chi)$. The constant of integration $x_0$ determines the center of the solution. For the symmetric configuration, the kink center is at the origin, so $x_0=0$. 

By using the form of $\chi=\chi(x)$, we can rewrite the first equation \eqref{foPkk} as $\phi^\prime = P_\phi -f(\chi(x))$, which explicitly depends on the spatial coordinate. The presence of the function $f(\chi(x))$ modifies the profile of $\phi(x)$ resulting in interesting configurations.   

For the solutions of the Eqs.~\eqref{fokk}, we can separate the energy density \eqref{rhoW} into two contributions, $\rho(x)=\rho_1(x) +\rho_2(x)$, where
\bes
\bal
\label{rhoP}\rho_1(x) &= P_\phi\,(P_\phi -f(\chi)),\\
\label{rhoC}\rho_2(x) &= Q_\chi^2.
\eal
\ees
The second contribution depends solely on the field $\chi$, with energy $E_2=Q(w_+)-Q(w_-)$, while the first one depends on both fields ($\phi$ and $\chi$), with energy $E_1=P(v_+)-P(v_-)$. Although the energy density $\rho_1$ depends on the fields $\phi$ and $\chi$, its energy depends only on the function $P(\phi)$ evaluated at the extrema of solution $\phi(x)$.

We can also study linear stability of the model. Indeed, by taking small fluctuations around the static fields in the form $\phi(x,t)=\phi(x)+\sum_n\eta_n(x)\cos(\omega_n t)$ and $\chi(x,t)=\chi(x)+\sum_n\xi_n(x)\cos(\omega_n t)$, one can substitute them in Eq.~\eqref{eom} to obtain
\bes\label{etaxistab}
\bal
&-\eta_n^{\prime\prime} -\frac{f_\chi\phi^\prime}{|\phi^\prime|}\xi_n^\prime  +V_{\phi\phi}\eta_n +\left(V_{\phi\chi} -\left(\frac{f_\chi\phi^\prime}{|\phi^\prime|}\right)^\prime\right)\xi_n = \omega_n^2\left(1 +\frac{f(\chi)}{|\phi^\prime|}\right)\eta_n,\\
&-\xi_n^{\prime\prime} +\frac{f_\chi\phi^\prime}{|\phi^\prime|}\eta_n^\prime +V_{\chi\phi}\eta_n +\big(V_{\chi\chi} +f_{\chi\chi}|\phi^\prime|\big)\xi_n = \omega_n^2\xi_n.
\eal
\ees
Despite the intricated form, we observe that, unlike the case of a single scalar field studied in \cite{cuscutonigor}, the cuscuton contributes to terms involving first derivatives of the perturbation $\eta^\prime_n$ and $\xi^\prime_n$. However, its contributions with terms involving second derivatives remain absent. This will be investigated in another work.

{In more general situations, the dynamical term $\p_\mu\phi\p^\mu\phi$ can vanish, leading to a divergence in the terms where this quantity is in the denominator, both in the equation of motion and in the energy-momentum tensor in Eqs.~\eqref{eom} and \eqref{Tmn}. This also occurs when studying the cuscuton term in cosmology, where the scalar field depends on the temporal coordinate $\phi=\phi(t)$. For this specific case, there is a cancellation due to other contributions that also vanish. In the case of static solutions, the behavior is similar. As we have seen, there is no divergence when considering fields dependent only on the spatial coordinate. One may wonder if the fluctuations $\eta_n(x)$ and $\xi_n(x)$ added in the expressions above Eq.~\eqref{etaxistab} lead to divergencies. We have considered them to be sufficiently small to avoid this issue. Despite that, in scenarios where temporal and spatial dependencies freely compete, such as in the study of kink collisions and the formation of defect networks in higher dimensions, this model needs to be further investigated.  One possibility is to introduce a generalization where, in a certain limit of a given parameter, we obtain the model studied here. With this, it would be possible to regularize and understand the behavior of possible divergences. This type of approach was used in Refs.~\cite{deSouzaDutra:2005cc,Simas:2017fjt} in the context of divergent solutions associated to vacuumless potentials.}

\section{Examples}\label{sec3}
In this section, we discuss some examples that allow us to obtain analytical solutions of the model introduced in this paper. For this purpose, we choose the following functions
\be
P(\phi)=\lambda \left(\phi - \frac13 \phi^3 \right) \quad\text{and}\quad Q(\chi)=\chi-\frac13 \chi^3,
\ee
with $\lambda>0$. As $P$ and $Q$ are odd functions, the models studied in this section present $Z_2 \times Z_2$ symmetry. The kink solution of the $\chi$ field can be obtained using Eq.~\eqref{eqC}
\be\label{solC}
\chi(x)=\tanh(x),
\ee
connecting $w_-=-1$ to $w_+=1$. The second contribution of the energy density \eqref{rhoC} is written as $\rho_2(x)=\sech^4(x)$, that can be integrated all over space to determine $E_2=4/3$. The presence of the term $f(\chi)$ leads to an explicit dependence on the spatial coordinate in Eq.~\eqref{foPkk}, so we get
\be
\phi^\prime = \lambda\left(1-\phi^2\right) -f(\tanh(x)) \label{fophi}.
\ee
To write the energy density $\rho_1(x)$ in Eq.~\eqref{rhoP}, we must solve the equation above. Of course that, in the uncoupled case ($f(\chi)=0$), the above equation has the solution
\be\label{solbeta0}
\phi(x)=\tanh(\lambda\,x),
\ee
whose asymptotic behavior is $\phi(\pm \infty)=\pm 1$. We take three specific functions that  preserve these boundary conditions, 
\bes  
\bal  
&f_I(\chi)=\beta\left(1-\chi^2\right), \label{typeI}
\\ 
&f_{II}(\chi) = \beta\,\chi^2\!\left(1 -\chi^2\right), \label{typeII} \\  
&f_{III} (\chi) = \beta\left(1 -a\chi^2\right)^2\left(1 -\chi^2\right). \label{typeIII}
\eal  
\ees
These functions correspond to type-I, type-II, and type-III models, where $\beta$ and $a$ are real parameters.  As seen, the energy is independent of the choice of the function $f(\chi)$ and we obtain $E_1=4\lambda/3$. However, the explicit expression for the energy density $\rho_1(x)$ depends on the solution $\phi(x)$, which we determine in the following subsections.

\subsection{Type-I models}\label{subsec1}
We start by considering the function $f_I(\chi)$ from Eq.~\eqref{typeI}. This function is proportional to $Q_\chi$ and, therefore, vanishes at $\chi=\pm 1$  and have maxima in $\chi=0$, with $f(0)=\beta$, reaching a maximum at $\chi=0$ with $f(0)=\beta$. The first-order equation \eqref{fophi} becomes
\be
\phi^\prime = \lambda\left(1-\phi^2\right) -\beta \, \sech^2(x),
\ee
where $\lambda$ and $\beta$ are the parameters that control the model. The kink solution for the above equation is
\be\label{sol1}
\phi(x) = \frac{1}{2\lambda}\left(\sqrt{1 +4\beta\lambda} +1\right)\tanh(x) +\frac{1}{2\lambda}\left(\sqrt{1 +4\beta\lambda} +1 -2\lambda\right)\left(\frac{P_+(x) +C_0Q_+(x)}{P_-(x) +C_0Q_-(x)}\right).
\ee
Here, $C_0$ is an integration constant that arises in the process and the functions $P_{\pm}(x)$ and $Q_{\pm}(x)$ are given by
\bes
\bal
P_{\pm}(x) = P_{\nu_{\pm}}^\lambda(\tanh(x)) \,\,\,\,\,\,\,\,{\rm and} \,\,\,\,\,\,\,\, Q_{\pm}(x) = Q_{\nu_{\pm}}^\lambda(\tanh(x)),
\eal
\ees
with
\be
\nu_{\pm} = \frac12\left(\sqrt{1 +4\beta\lambda}\, \pm1\right).
\ee
$P_\nu^\mu(z)$ and $Q_\nu^\mu(z)$ are the associated Legendre functions of the first and second kind with argument $z$ and parameters $\mu$ and $\nu$. We choose the constant of integration $C_0$, such that $\phi(0)=0$. This leads us 
\be
C_0 = -\frac{Q_{\nu_+}^\lambda(0)}{P_{\nu_+}^\lambda(0)}.
\ee
For $\beta=0$, the solution in Eq.~\eqref{sol1} reduces to the kink solution given in Eq.~\eqref{solbeta0}.

\begin{figure}[htb!]
\centering
\includegraphics[width=5.9cm,trim={0.6cm 0.7cm 0 0},clip]{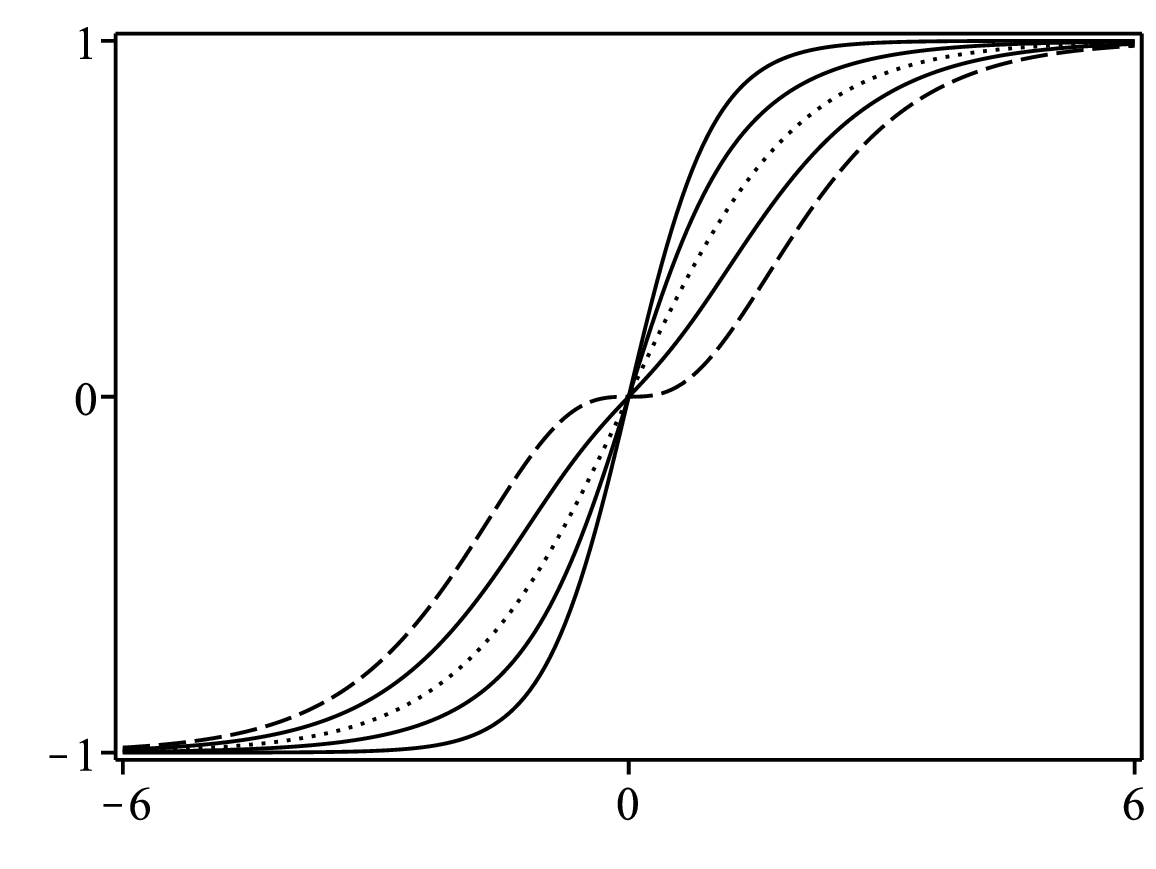}
\includegraphics[width=5.9cm,trim={0.6cm 0.7cm 0 0},clip]{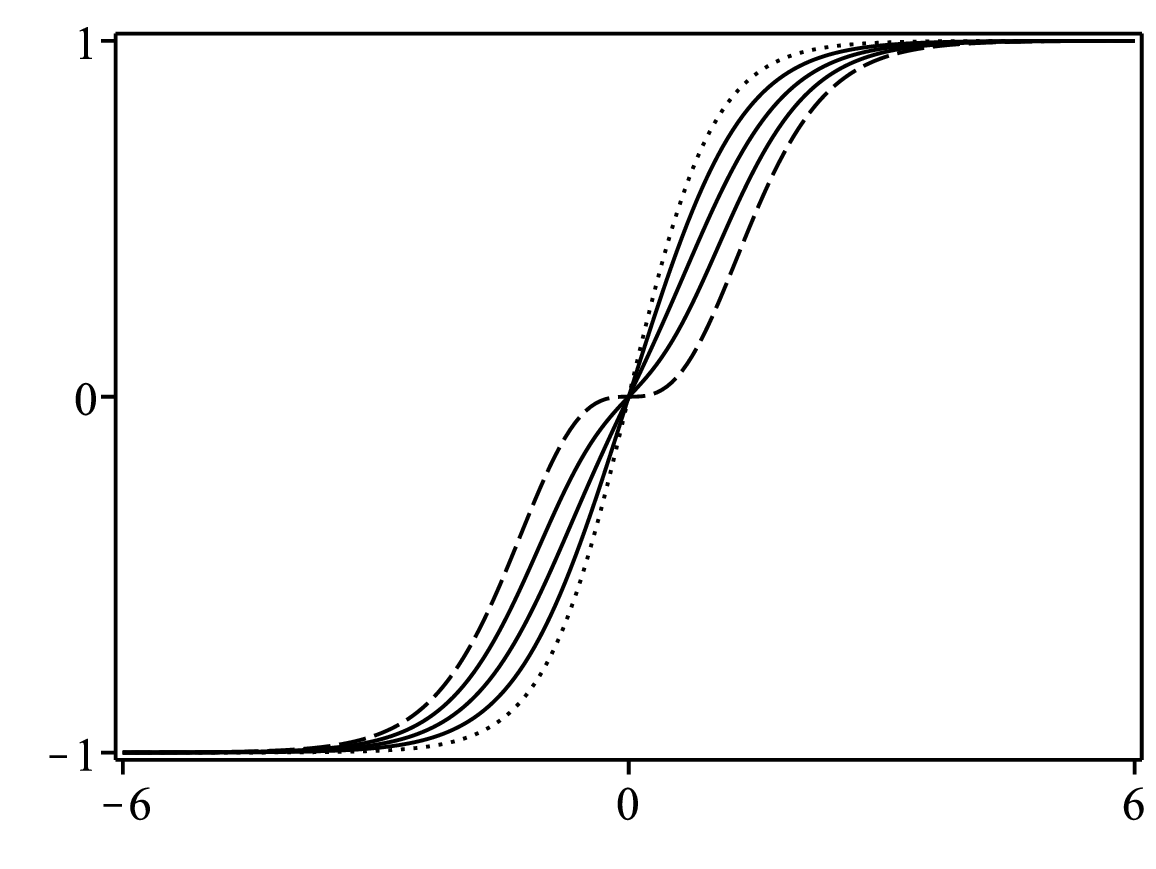}
\includegraphics[width=5.9cm,trim={0.6cm 0.7cm 0 0},clip]{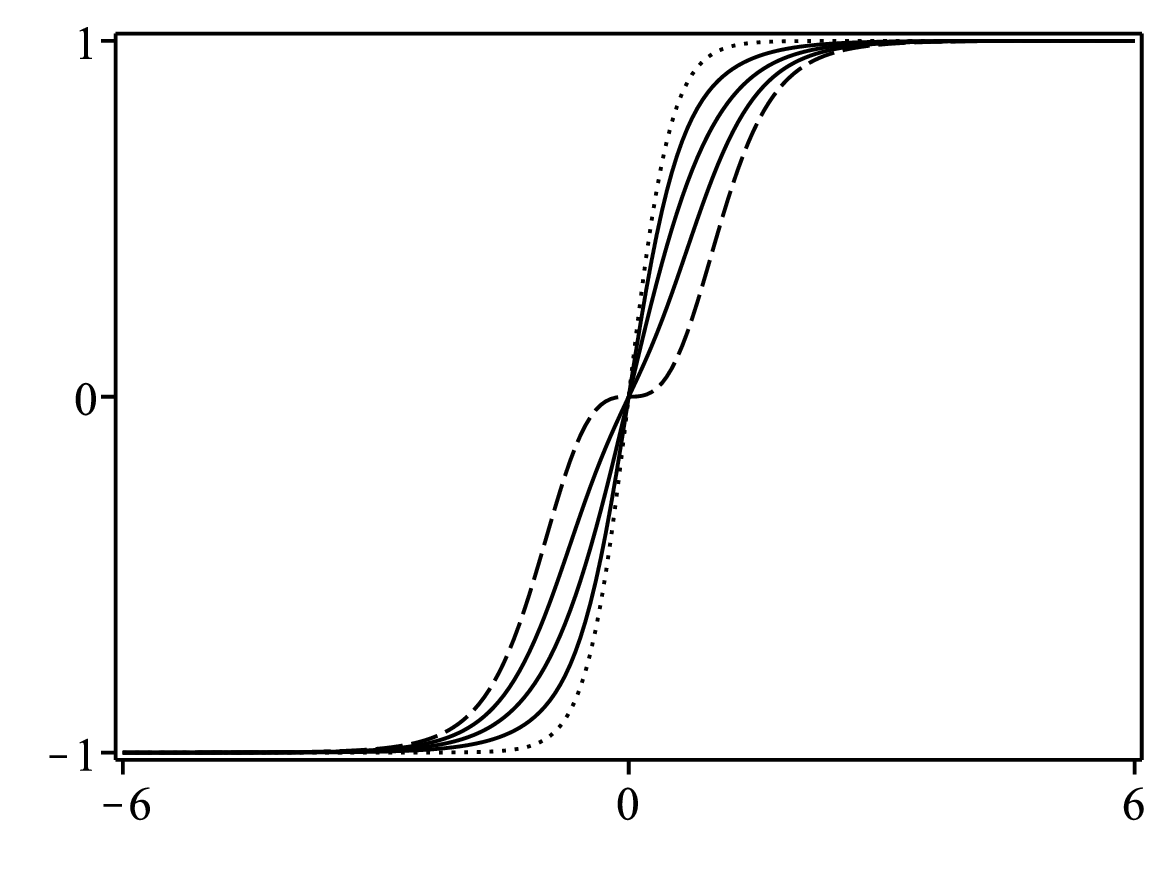}
\caption{The solution $\phi(x)$ in Eq.~\eqref{sol1}. In left panel, it is displayed for $\lambda=1/2$, and $\beta=0$ (dotted line), $-1/2,-1/4,1/4$ (solid lines) and $1/2$ (dashed line). In the middle panel, it is shown for $\lambda=1$, and  $\beta=0$ (dotted line), $1/4,1/2,3/4$ (solid lines) and $1$ (dashed line). In the right panel, it is depicted for $\lambda=2$, and $\beta=0$ (dotted line), $1/2,1,3/2$ (solid lines) and $2$ (dashed line).}
\label{fig1}
\end{figure}

According to the Eq.~\eqref{sol1}, for $\lambda\geq1$, the range of the parameter $\beta$ is $\beta \in [0,\lambda]$, that is always non-negative; while for $\lambda<1$ it is $\beta \in [\lambda-1,\lambda]$, allowing negative values. In particular, when $\beta=\lambda-1$, the solution for $\phi(x)$ is identical to the solution for the field $\chi$, given by Eq.~\eqref{solC}. For the upper limit, $\beta=\lambda$, there is a plateau in the solution at $x=0$; for this reason, we denote $\beta_c=\lambda$ as critical value. To analyze the behavior of Eq.~\eqref{sol1} near the origin, we can use a power series expansion, which yields
\be
\phi(x) = (\beta_c-\beta)\,x +\left(\beta -\beta_c(\beta_c-\beta)^2\right)\frac{x^3}{3} +\mathcal{O}\!\left(x^5\right).
\ee
For the critical value $\beta=\beta_c$, the linear contribution vanishes and the leading term becomes $\phi(x\approx 0)\approx (\beta_c/3)\,x^3$. This behavior  is shown in Fig.~\ref{fig1}, where we plot the solutions \eqref{sol1} for  $\lambda=1/2$ and $1$, and some values of $\beta$. The parameter $\beta$ smoothly modifies the slope of the solutions at the origin: as it approaches its critical value ($\beta_c$), the slope decreases until zero. The thickness of these configurations depends on $\lambda^{-1}$, as usual. 
 
\begin{figure}[htb!]
\centering
\includegraphics[width=5.9cm,trim={0.6cm 0.7cm 0 0},clip]{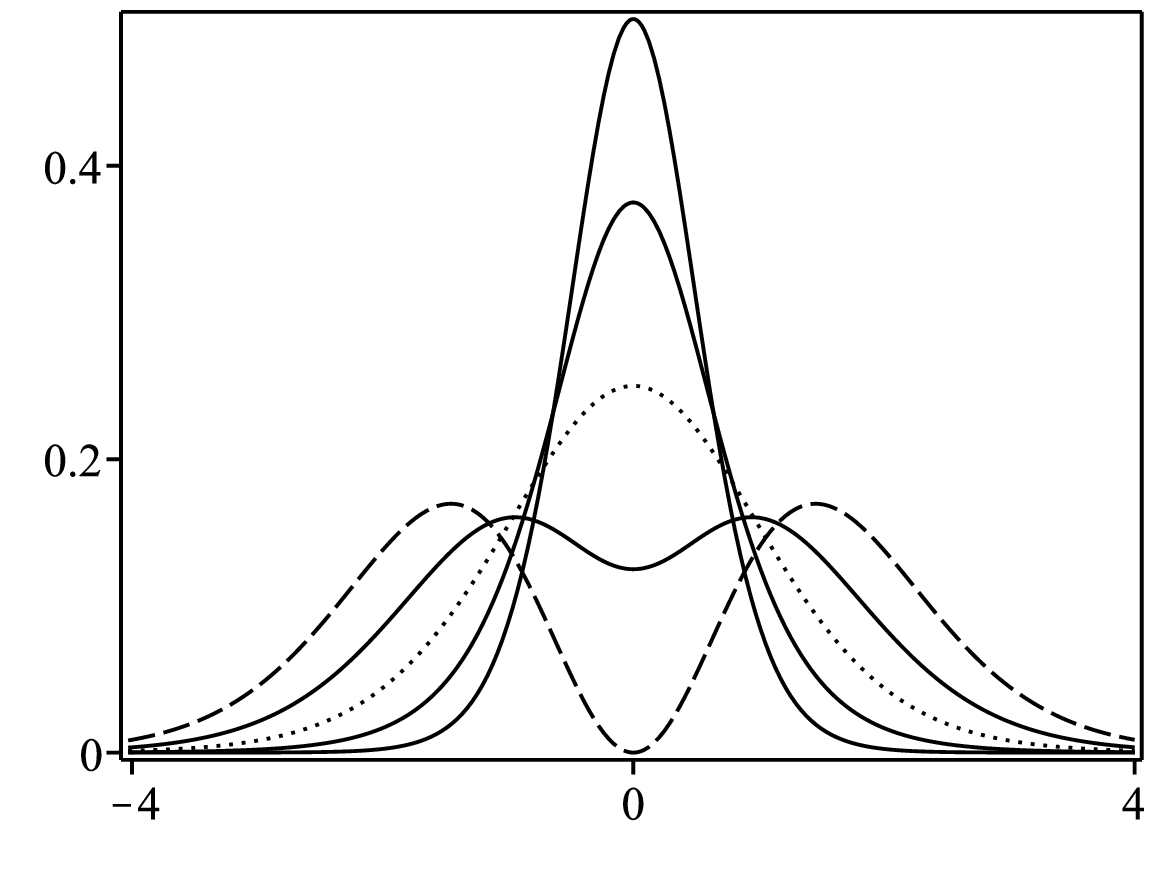}
\includegraphics[width=5.9cm,trim={0.6cm 0.7cm 0 0},clip]{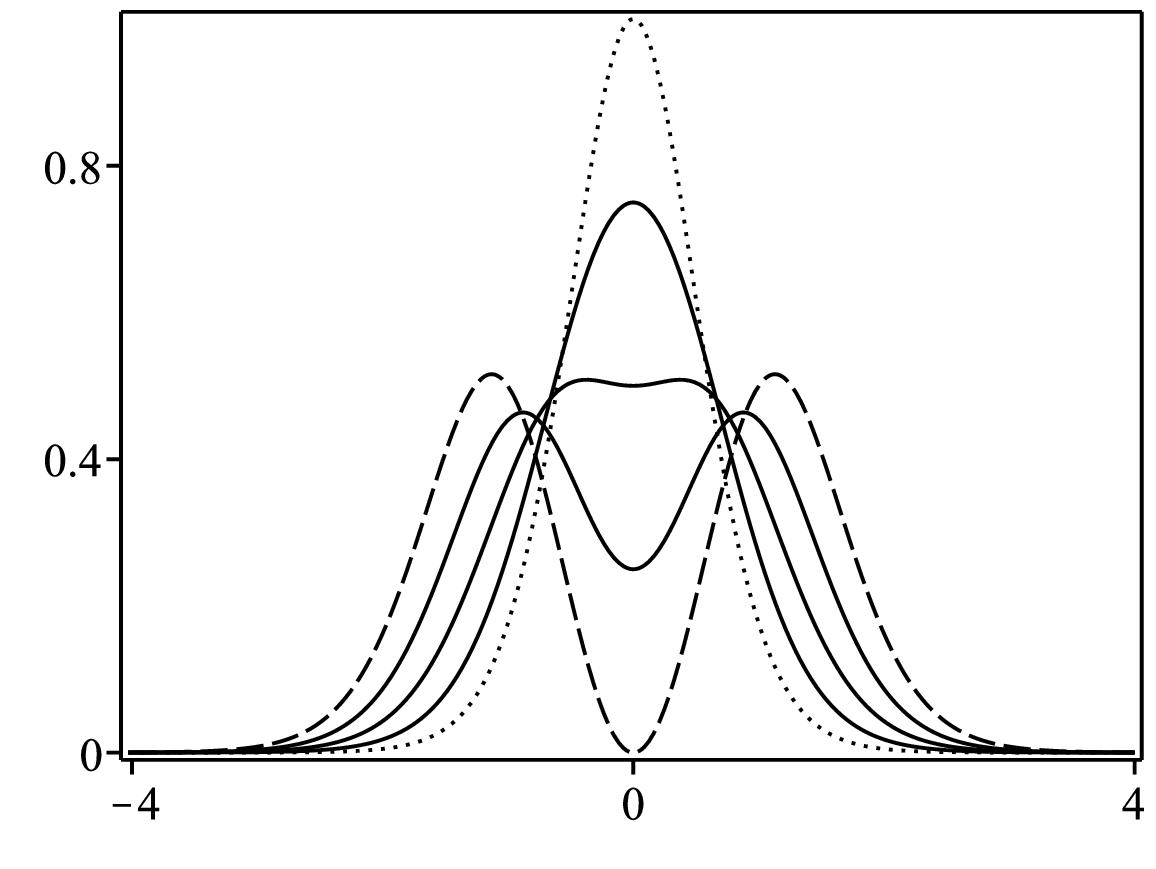}
\includegraphics[width=5.9cm,trim={0.6cm 0.7cm 0 0},clip]{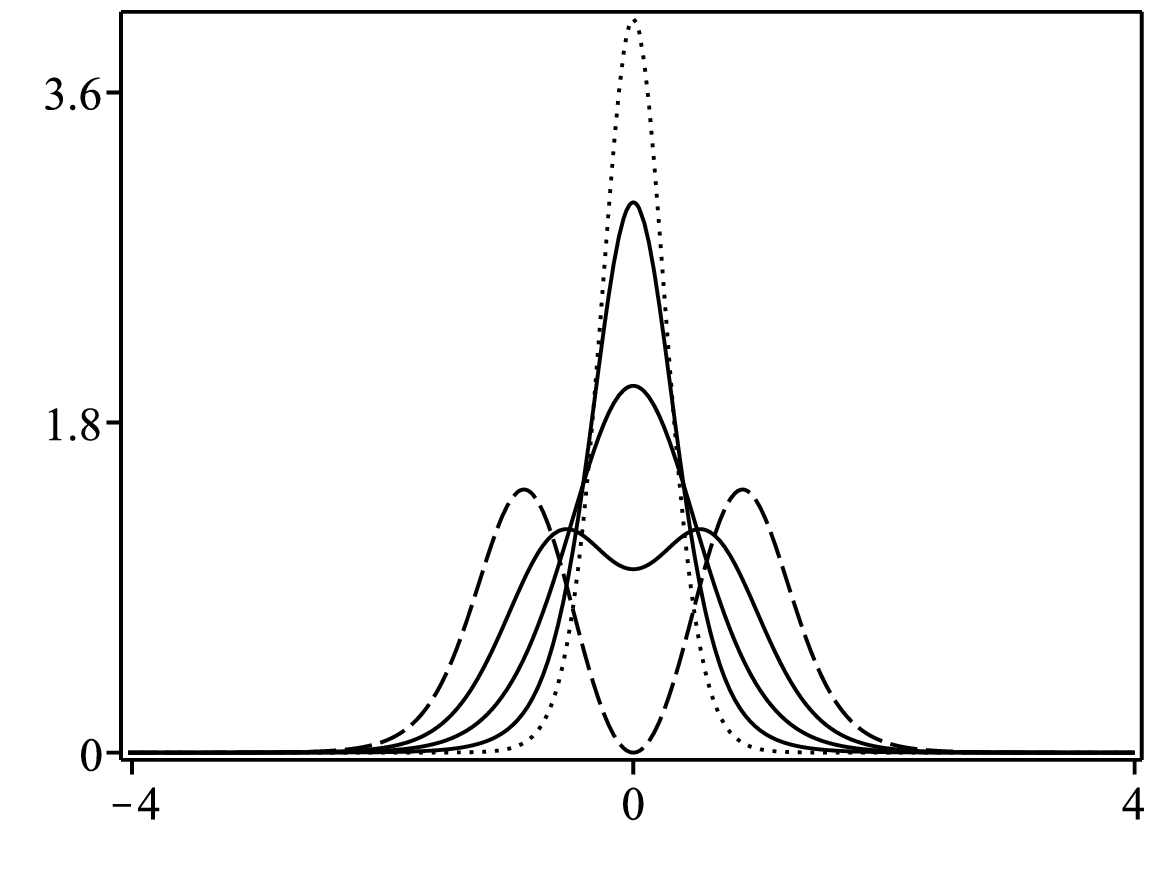}
\caption{Contribution $\rho_1(x)$ to the energy density \eqref{rhoP} for the solutions \eqref{sol1} with the same line styles and values of the parameters $\lambda$ and $\beta$ in Fig.~\ref{fig1}.}
\label{fig2}
\end{figure}

An analytical expression for the contribution $\rho_1(x)$ to the energy density \eqref{rhoP} can be obtained, but it is cumbersome and therefore was omitted here. We can investigate this quantity near the origin, similarly to what was done for the solution, 
\be
\rho_1(x) = \beta_c(\beta_c-\beta) +\beta_c\left(\beta -(2\beta_c -\beta)(\beta_c-\beta)^2\right)x^2 +\mathcal{O}\!\left(x^4\right).
\ee
Note that the intensity of the energy density at the origin is $\rho_1(0)=\beta_c(\beta_c-\beta)$, which can be a maximum or a minimum depending on the values of $\lambda$ and $\beta$.  For the critical value $\beta=\beta_c$, the quantity $\rho_1(0)$ vanishes, and the energy density is symmetrically divided into two portions. This behavior is illustrated in Fig.~\ref{fig2}, where we display this quantity for the same values of $\lambda$ and $\beta$ as in the previous figure.

\subsection{Type-II models}
\begin{figure}[htb!]
\centering
\includegraphics[width=5.9cm,trim={0.6cm 0.7cm 0 0},clip]{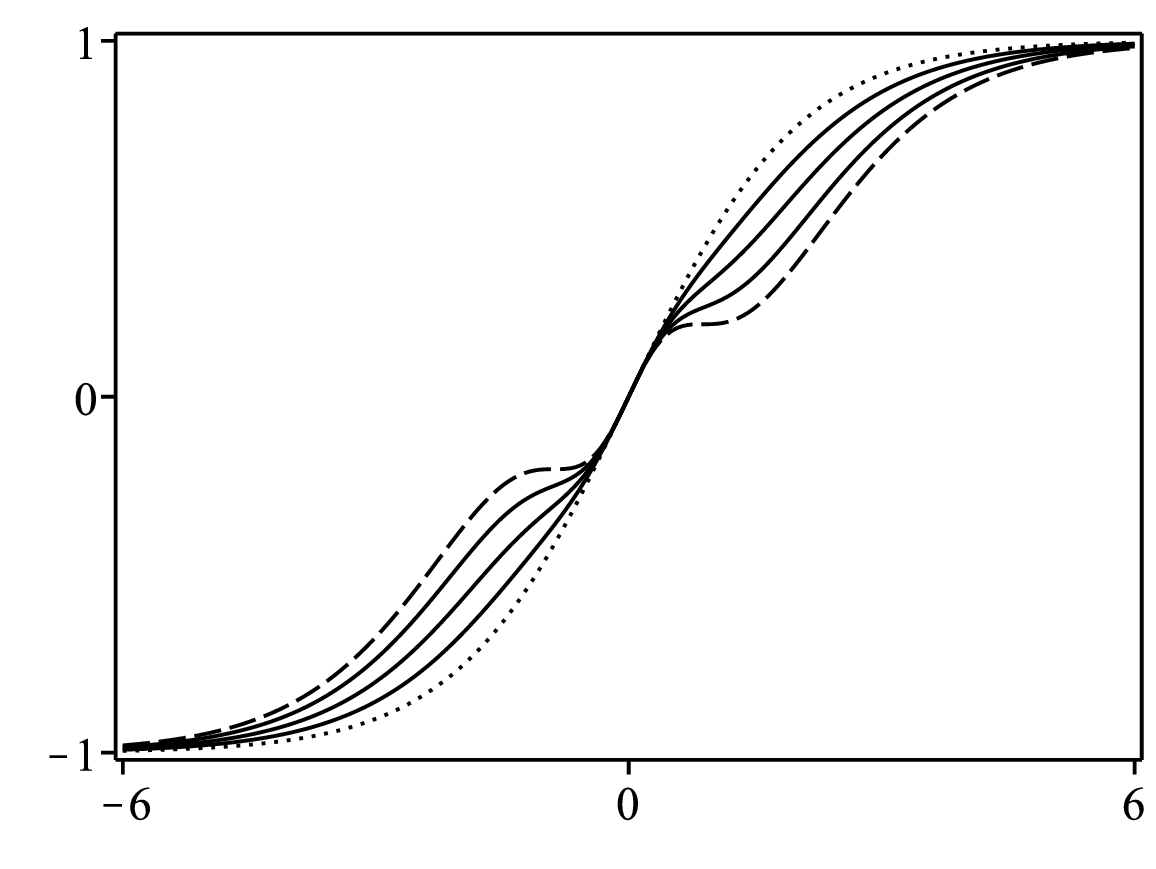}
\includegraphics[width=5.9cm,trim={0.6cm 0.7cm 0 0},clip]{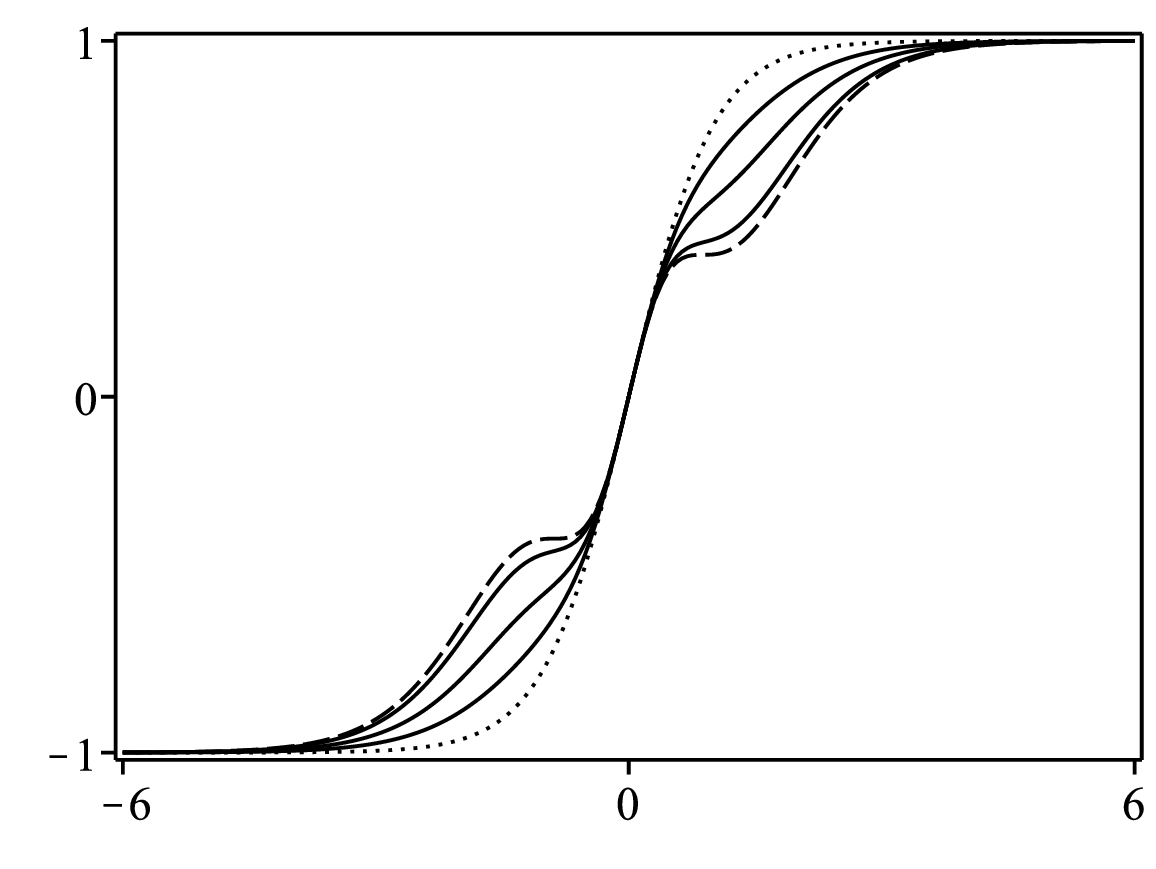}
\includegraphics[width=5.9cm,trim={0.6cm 0.7cm 0 0},clip]{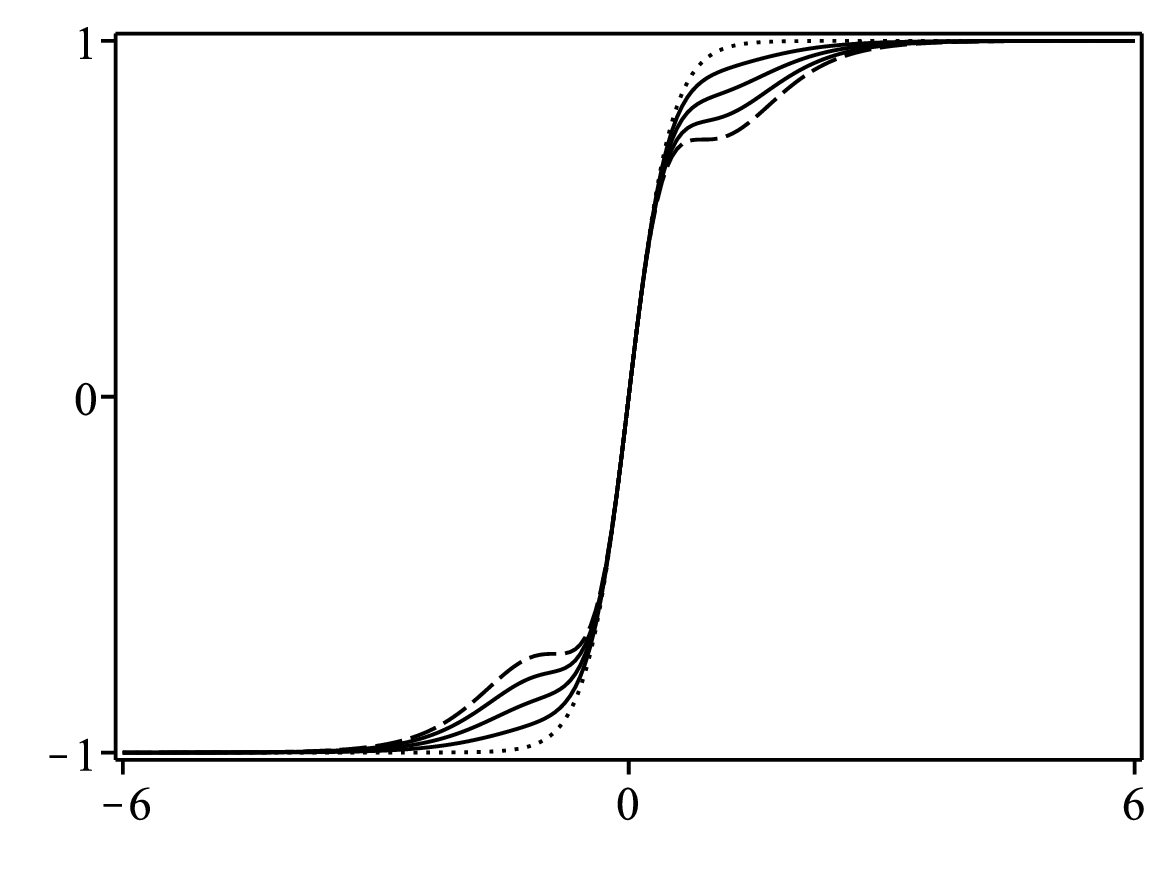}
\caption{The solution $\phi(x)$ in Eq.~\eqref{sol2}. In the left panel, it is depicted for $\lambda=1/2$, and $\beta=0$ (dotted line), $1/2,1,3/2$ (solid lines) and $=1.9170$ (dashed line). In the middle panel, it is displayed for $\lambda=1$, $\beta=0$ (dotted line), $1,2,3$ (solid lines) and $3.3639$ (dashed line). The right panel, it is shown for $\lambda=2$, $\beta=0$ (dotted line), $1,2,3$ (solid lines) and $3.8227$ (dashed line).}
\label{fig3}
\end{figure}

Now we consider novel configurations that arise with the function $f_{II}(\phi)$ given by Eq.~\eqref{typeII}. In this model, we take $\beta\geq0$. The first order equation \eqref{fokk} becomes
\be
\phi^\prime = \lambda\left(1-\phi^2\right) -\beta \, \sech^2(x)\tanh^2(x).
\ee
For $\beta=0$, the solution is given by \eqref{solbeta0}, and for $\beta>0$, it is given by
\be\label{sol2}
\phi(x) = \frac{d}{dx}\left(\ln^\frac{1}{\lambda}\left(\sech^\lambda(x)\,\text{HeunC}\!\left(0,-\frac12,\lambda,-\frac{\beta\lambda}{4},\frac{\lambda^2+1}{4};\tanh^2(x)\right)\right)\right),
\ee
where $\text{HeunC}(\alpha,\gamma,\delta,\epsilon,\eta;z)$ are the confluent Heun functions with the parameters $\alpha,\gamma,\delta,\epsilon,\eta$ and argument $z$. We fix the constant of integration to obtain $\phi(0)=0$. The parameters $\lambda$ and $\beta$ control the behavior of the solution. As in the previous model, the thickness of the solution is controlled by $\lambda$, and $\beta$ modifies the slope at points outside the origin. Instead, it changes the slope at symmetrical points about the origin. There is a maximum value $\beta$, which we define as $\beta_c$. For this value, the solution has null derivative at two points, $x_\pm\approx\pm 0.8813$, leading to the presence of two plateaus. In Fig.~\ref{fig3} we can see this behavior for $\lambda=1/2,1$ and $2$, and some values of $\beta$. As $\beta$ increases, the inclination of the solution decreases until two plateaus arise at $x_\pm$. For $\lambda=1/2,1$ and $2$ the critical values of $\beta$ are $\beta_c=1.9170,3.3639$ and $3.8227$, respectively. It is noteworthy that the parameter $\lambda$, in addition to modify the thickness of the solutions, shifts the value of the $\phi(x_\pm)$, which has zero derivative. 

\begin{figure}[htb!]
\centering
\includegraphics[width=5.9cm,trim={0.6cm 0.7cm 0 0},clip]{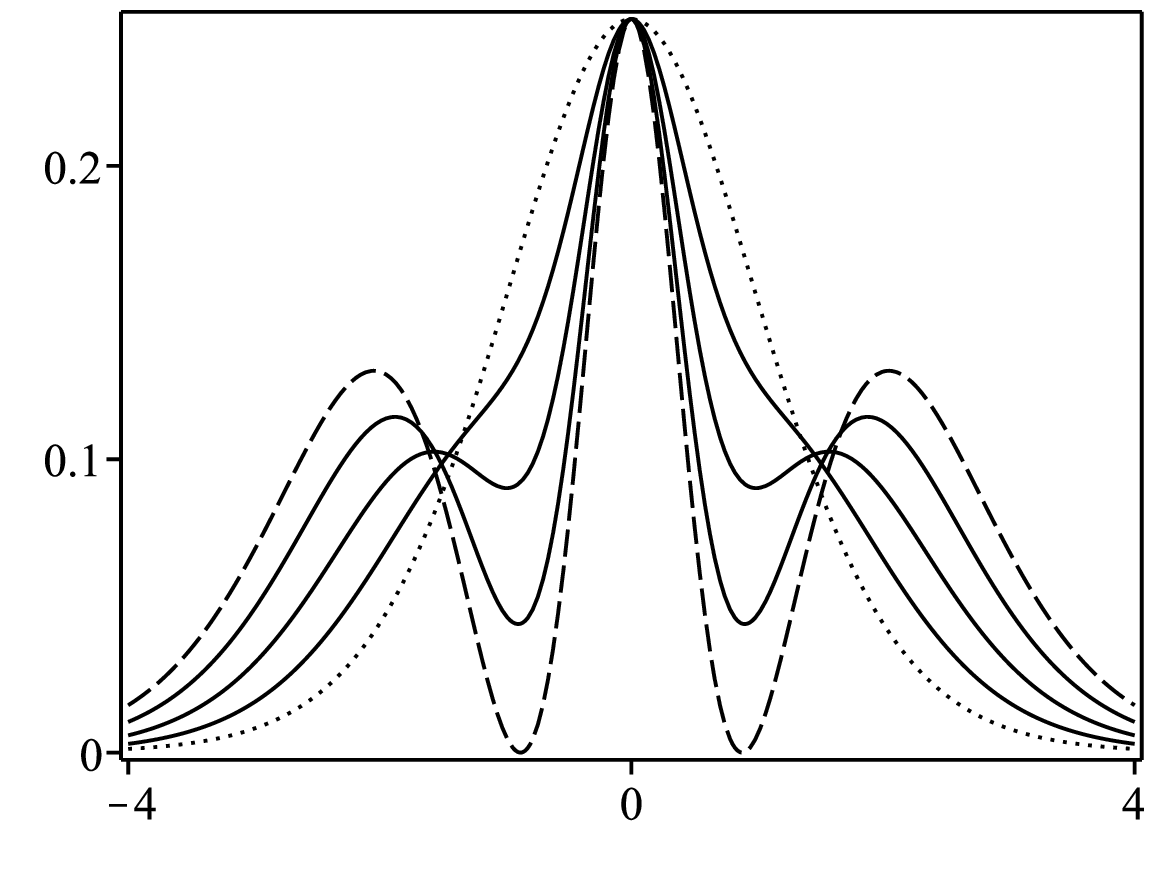}
\includegraphics[width=5.9cm,trim={0.6cm 0.7cm 0 0},clip]{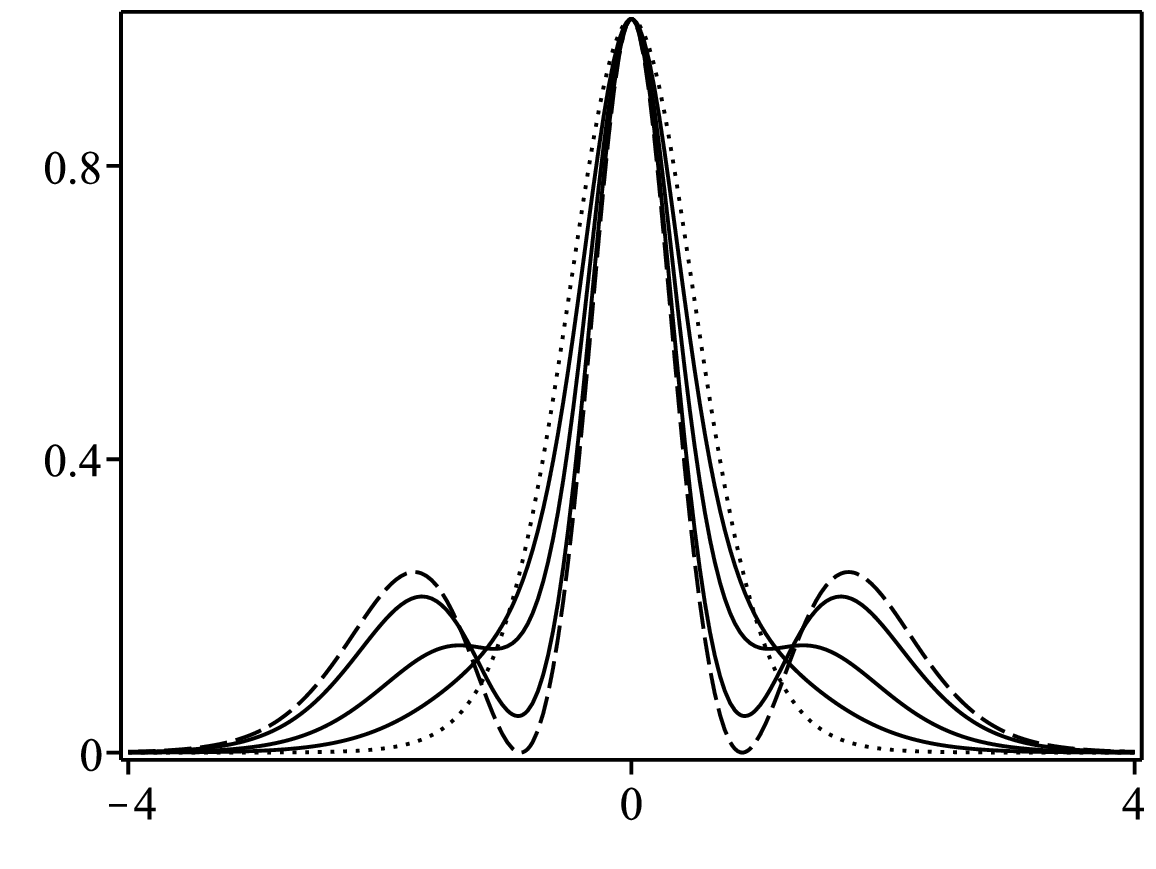}
\includegraphics[width=5.9cm,trim={0.6cm 0.7cm 0 0},clip]{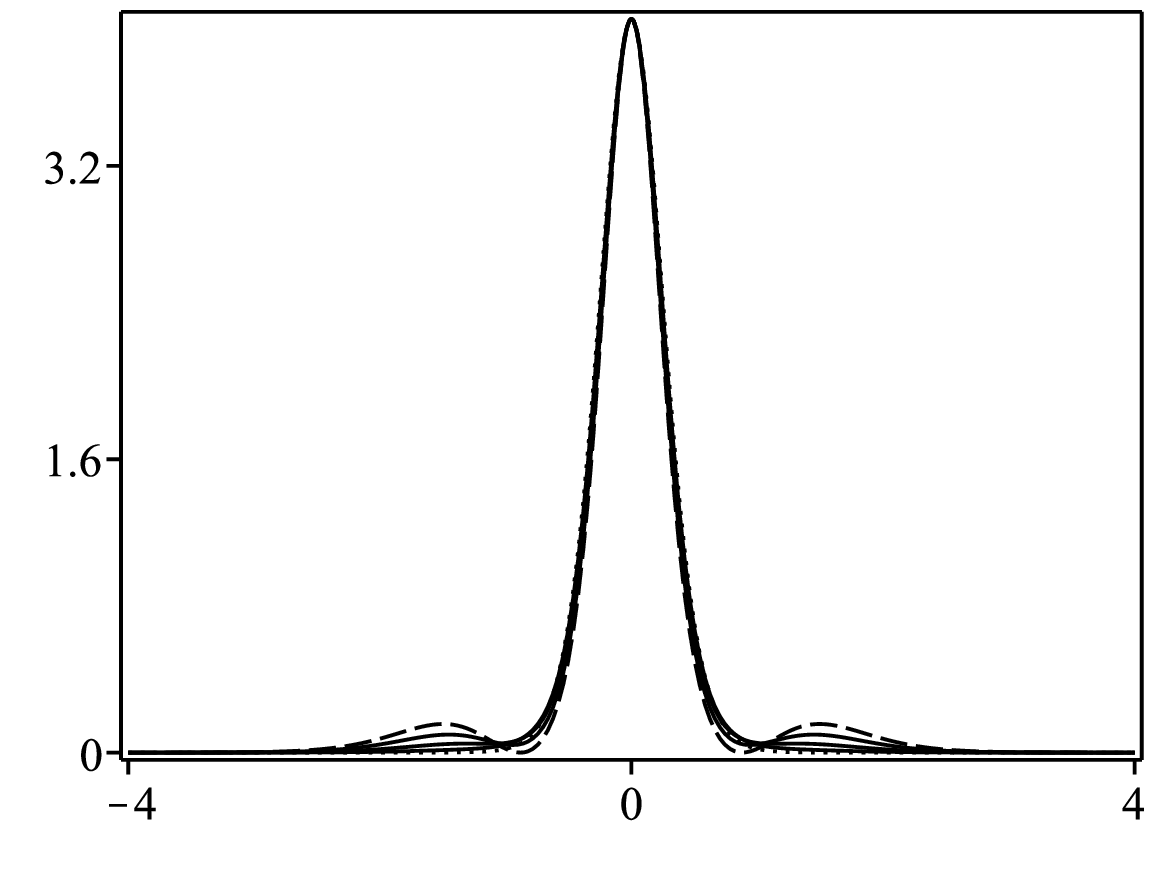}
\caption{Contribution $\rho_1(x)$ of the energy density \eqref{rhoP} for the solutions \eqref{sol2} with the same line styles and values of the parameters $\lambda$ and $\beta$ in Fig.~\ref{fig3}.}
\label{fig4}
\end{figure}

The demeanor of the contribution $\rho_1(x)$ to the energy density \eqref{rhoP} for this solutions is shown in Fig.~\ref{fig4} for several values of $\lambda$ and $\beta$. One can see that always has a maximum at the center with intensity $\rho_1(0)=\lambda^2$. As $\beta$ increase, local minima and maxima appear. For $\beta=\beta_c$, the minima turn global minima at $x_\pm$, with $\rho_1(x_\pm)=0$. At this critical value, the contribution $\rho_1(x)$ splits in one central and two outer portions, whose ratio between then depends on $\lambda$. as can see in figure.

\subsection{Type-III models}
\begin{figure}[t!]
\centering
\includegraphics[width=5.9cm,trim={0.6cm 0.7cm 0 0},clip]{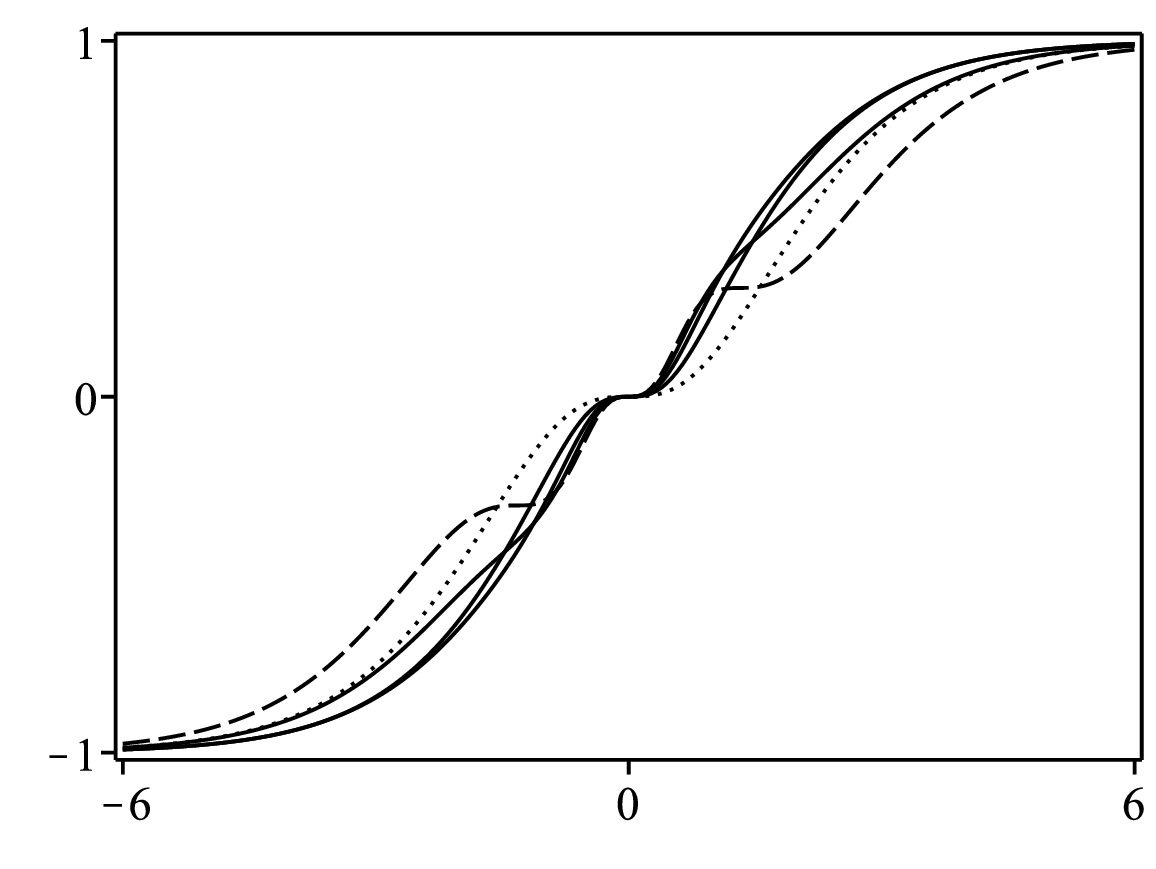}
\includegraphics[width=5.9cm,trim={0.6cm 0.7cm 0 0},clip]{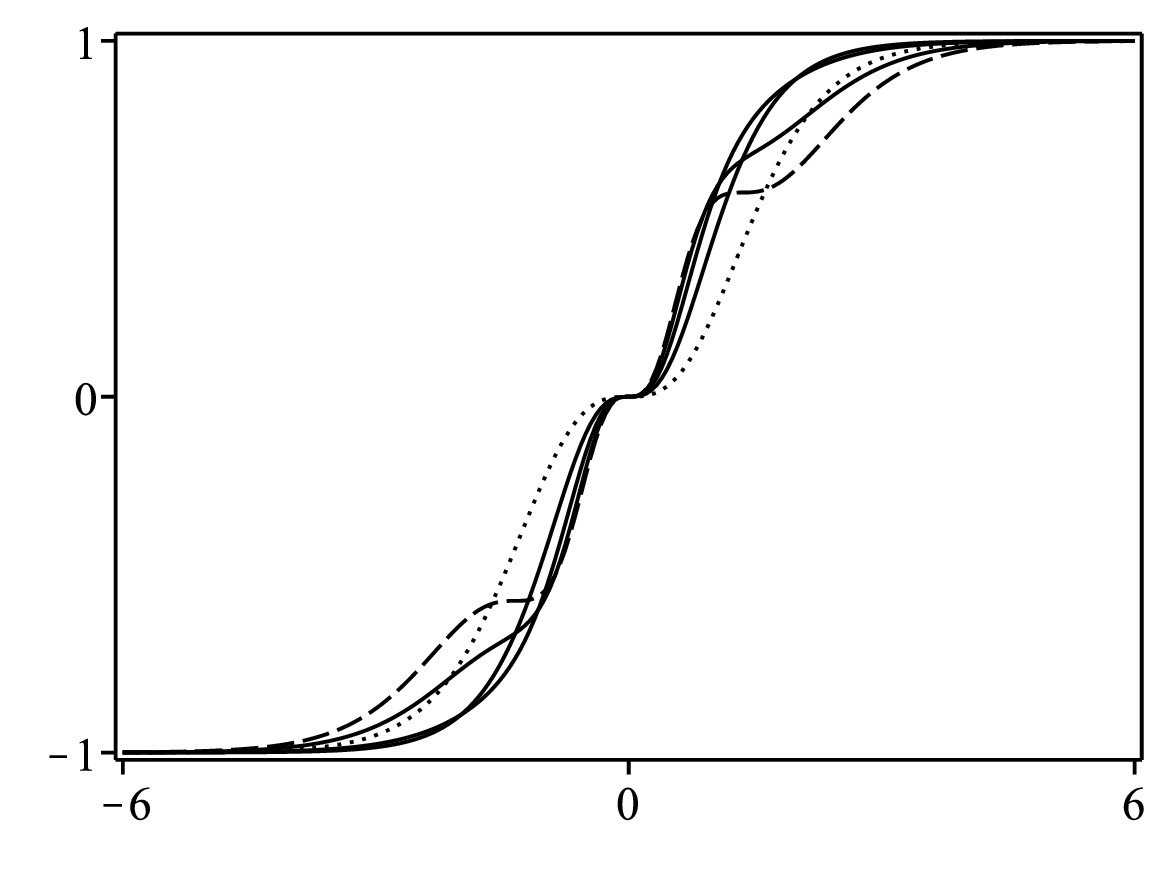}
\includegraphics[width=5.9cm,trim={0.6cm 0.7cm 0 0},clip]{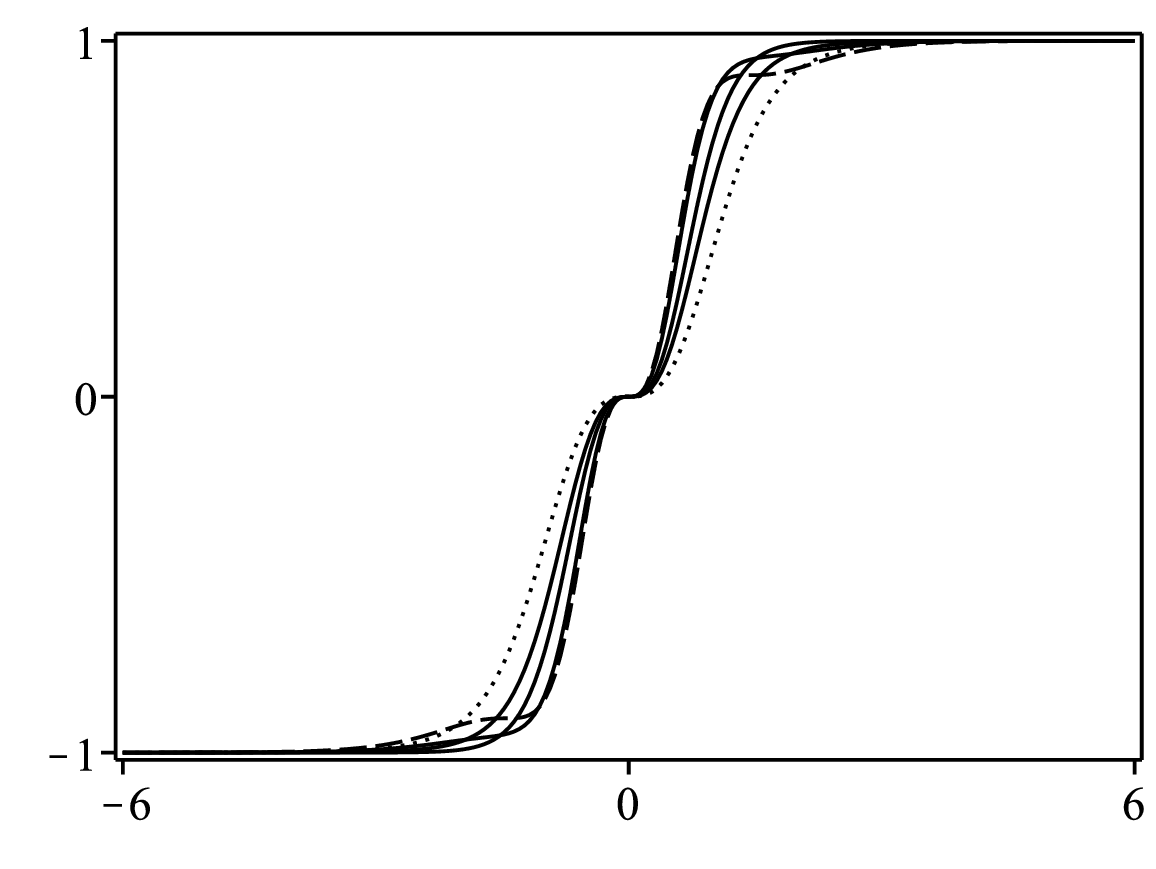}
\caption{The solutions \eqref{sol3} for $\beta=\lambda$ and some values of $\lambda$ and $a$. In the left panel, it is shown for $\lambda=1/2$, and $a=0$ (dotted line), $1,2,3$ (solid lines) and $3.8723$ (dashed line). In the middle panel, it is depicted for $\lambda=1$, $a=0$ (dotted line), $1,2,3$ (solid lines) and $3.5156$ (dashed line). In the right panel, it is displayed for $\lambda=2$, $a=0$ (dotted line), $1/2,1,2$ (solid lines) and $2.4471$ (dashed line).}
\label{fig5}
\end{figure}

We now consider another class of models described by the function $f_{III}(\chi)$ in the Eq.~\eqref{typeIII}. The first order equation \eqref{fokk} becomes 
\be
\phi^\prime = \lambda\left(1-\phi^2\right) -\beta\, \sech^2(x)(1-a\tanh^2(x))^2,
\ee
with $\beta \in [0,\beta_c]$ and $a \in [-1/2, a_c]$, where $\beta_c=\lambda$ and $a_c$ is determined for specific values of $\lambda$ and $\beta$. When $a=0$, the model reduces to the type-I model discussed in Sec.~\ref{subsec1}.  The solution is given by
\be\label{sol3}
\begin{aligned}
\phi(x) &= \frac{d}{dx}\left(\ln^\frac{1}{\lambda}\left(\sech^\lambda(x)\,\text{HeunC}\!\left(a\sqrt{\beta\lambda},-\frac12,\lambda,-\frac{\beta\lambda a(a-2)}{4},\frac{1+\lambda^2-\beta\lambda}{4};\tanh^2(x)\right)\right)\right)\\
    &+a\sqrt{\frac{\beta}{\lambda}}\,\sech^2(x)\tanh(x),
\end{aligned}
\ee
where we have fixed the constant of integration to obtain $\phi(0)=0$. The two parameters provide new features for the kink solution. As in the type-I model, $\beta$ modifies the inclination at the origin, and becomes null at its upper value, $\beta=\beta_c$. This behavior can be seen by analyzing the solution around $x=0$,
\be
\phi(x) = A\,x +B\,x^3 +C\,x^5 +\mathcal{O}\!\left(x^7\right),
\ee
where
$A = \beta_c-\beta,
B = -\beta_c A^2/3+\beta(1+2a)/3$ and $C = -2\beta_cAB/5 -\beta\left(2+10a+3a^2\right)\!/15.$ In addition with respect to the type-I model, the parameter $a$ also contributes to modifying this feature. For $\beta=\beta_c$ and $a\neq-1/2$ the linear contribution disappears ($A=0$), giving rise to a plateau where the cubic dependence is dominant, similar to the aforementioned model. When $\beta=\beta_c$ and $a=-1/2$, the linear and cubic terms vanishes ($A=0, B=0$), making the plateau wider; for this case, we have $\phi(x\approx 0)\approx (3\lambda/20)\,x^5$. Besides contributing to the behavior at $x=0$, the parameter $a$ also plays a role in modifying the slope of the solution at symmetrical points outside the center ($x\neq 0$). When $a=a_c$, the solution gets inflection points with null derivative, leading to two additional plateaus. To demonstrate this new feature, we plot the solution \eqref{sol3} in Fig.~\ref{fig5}, where we take $\beta=\beta_c$, $\lambda=1/2,1$ and $2$, and some values of $a$, including $a_c=3.8723,3.5156$ and $2.4471$. For these choices, we can see that the solution with one plateau gets two additional plateaus, resulting a three-plateau configuration for $a=a_c$. As we increase $\lambda$, the symmetric plateaus get closer to the tail of the solution.

\begin{figure}[htb!]
\centering
\includegraphics[width=5.9cm,trim={0.6cm 0.7cm 0 0},clip]{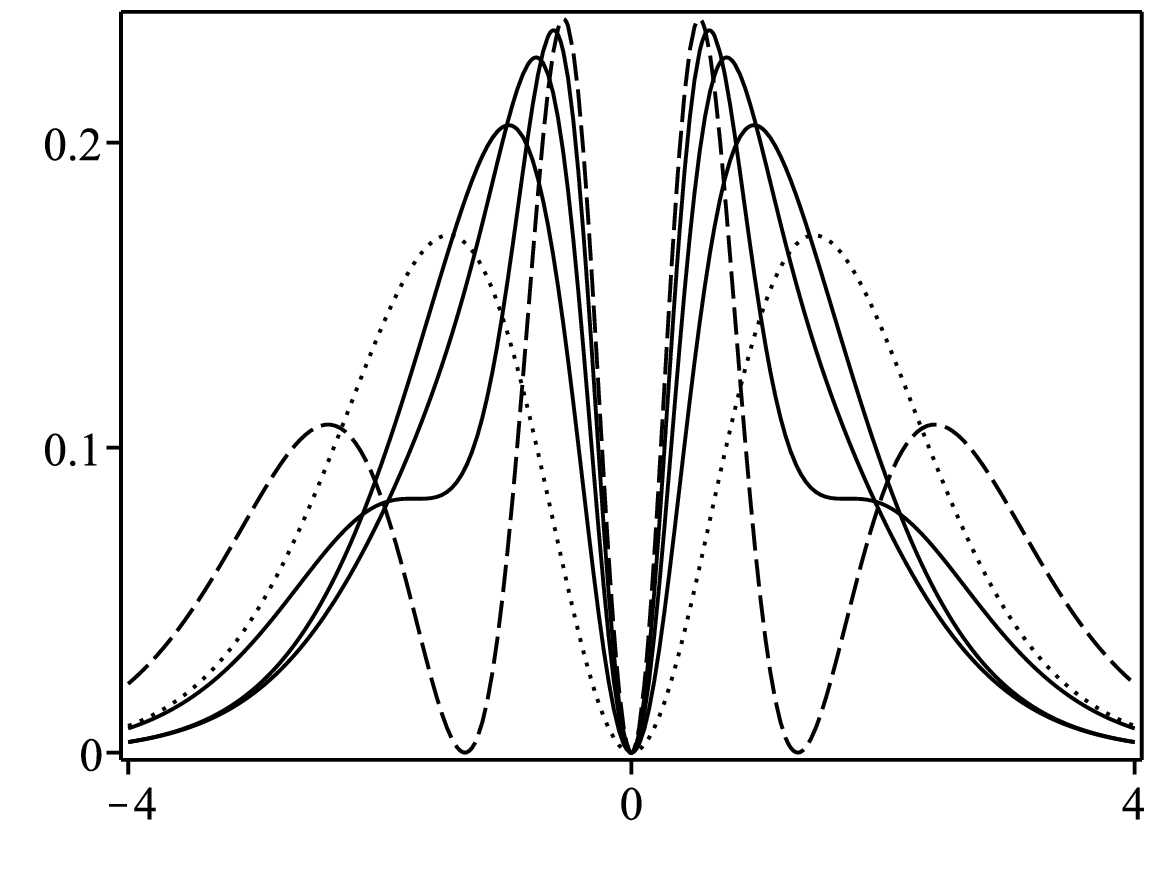}
\includegraphics[width=5.9cm,trim={0.6cm 0.7cm 0 0},clip]{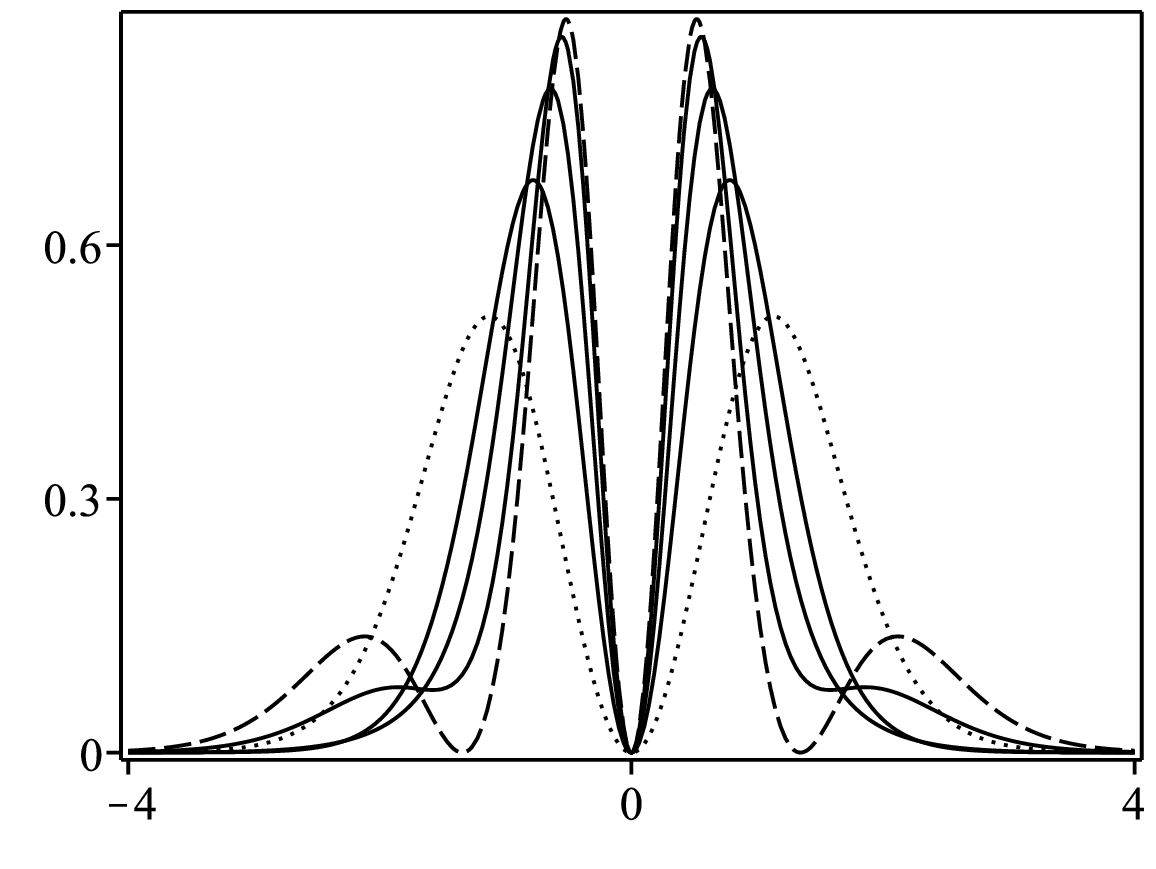}
\includegraphics[width=5.9cm,trim={0.6cm 0.7cm 0 0},clip]{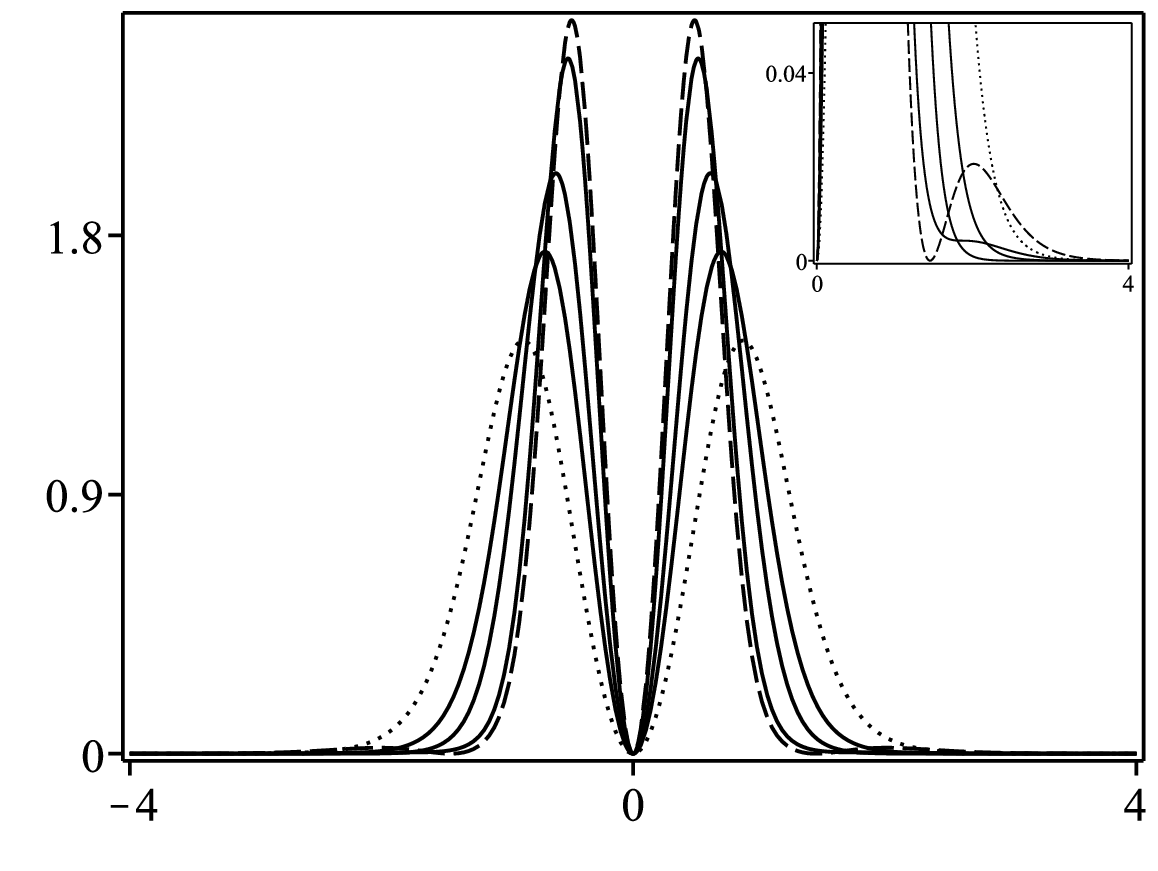}
\caption{Contribution $\rho_1(x)$ to the energy density \eqref{rhoP} for the solutions \eqref{sol1} with the same line styles and values of the parameters $\lambda$, $\beta$ and $a$ in Fig.~\ref{fig5}. In the right panel, in the inset, we enlarged the tail to better visualize its peripheral portion.}
\label{fig6}
\end{figure}

Analogously to the previous examples, we omit the expression of the energy density $\rho_1(x)$ given by Eq.~\eqref{rhoP}, but we depict it in Fig.~\ref{fig6} for same choices of parameters  utilized in Fig.~\ref{fig5}. An analysis close to the origin reveals that
\be
\rho_1(x) = \beta_cA +\beta_c\left(3B -A^2\right)x^2 +5\beta_c\left(C -A^2B\right)x^4 +\mathcal{O}\!\left(x^6\right),
\ee
where the dominant contributions depend on the chosen parameters. It is clear that for $\beta=\beta_c$ and $a\neq -1/2$ the intensity at the origin is null, so $\rho_1(x\approx 0)\approx \beta_c^2(1+2a)\,x^2$. Finally, in the additional case $\beta=\beta_c$ and $a=-1/2$, we have $\rho_1(x\approx 0)\approx (3\beta_c^2/4)\,x^4$. Notice that besides the hole in the center, there are also two symmetric others. As it was seen, this happens when we take $\beta=\beta_c$ and $a=a_c$. In this situation, $\rho_1(x)$ splits in four portions whose ratios depend on $\lambda$. For example, for $\lambda=2$, the contribution of the peripheral pieces is much smaller than that of the others.

\section{End Comments}\label{sec4}
 In this work we have investigated a class of models of two scalar fields in the presence of a new coupling inspired by cuscuton dynamics. Unlike the case of a single field model investigated in Ref.~\cite{cuscutonigor}, here, this addition contributes in the equations of motion for the static case, without introducing second-derivative terms of the field. To find the solutions, one must solve the equations of motion. Since they are of second order with couplings between the fields, we have developed a formalism by introducing smooth functions that lead to first-order equations for kink solutions, providing a simple and direct way to obtain their energy. We introduced three classes of examples from which we obtain analytical solutions with distinct behaviors

Using the aforementioned procedure, we have developed a first-order equation for the second field $\chi$ that only depends on $Q(\chi)$. Then, we have fixed the same $\chi$ solution for the all examples and focused our attention on how the field $\phi$ behaves for specific choices of $f(\chi)$. The right-hand side of the first-order equation \eqref{foPkk} contains two terms, with one of them depending on both $\phi$ and $\chi$ and competing with each other. We have shown that this new interaction leads to meaningful modulations in the kink profile. With appropriate choices for the parameters, it is possible to control the slope of the $\phi$ field and the internal structure of its energy density, which has plateaus and null intensity, respectively, for the critical situations.

The profile of the solutions described in the present work reminders us of the ones found in Ref.~\cite{liao,Sebastian:2023dds}, where the authors have dealt with the problem through a coupling in the usual dynamics term. This coupling leads to first order equations that differ from those presented here. In that work, they also investigated other models, which lead to asymmetric configurations that have not been considered here. An interesting direction for future research would be to explore this possibility.  Other perspectives include the study of the model in the presence of gravity, in the five-dimensional braneworld scenario, with a single extra dimension \cite{brana1,brana2,brana3,cuscutondouglas}. Another interesting issue concerns extending the above results for relativistic field theories that support vortex solutions, where a complex scalar field is coupled with a gauge field under the action of the local $U(1)$ gauge symmetry \cite{vortexM1,vortexM2,vortexCS1,vortexCS2}. We hope to report on these and other related issues in the near future.

\acknowledgements{We would like to thank D. Bazeia and M.A. Marques for the discussions that have contributed to this work. We acknowledge the financial support from the Brazilian agencies Conselho Nacional de Desenvolvimento Cient\'ifico e Tecnol\'ogico (CNPq), grant No. 306504/2018-9 (RM), and Para\'iba State Research Foundation (FAPESQ-PB), grant No. 0003/2019 (RM).}

\end{document}